# A Grounded Theory Based Approach to Characterize Software Attack Surfaces

Sara Moshtari
Rochester Institute of Technology
Rochester, NY, USA
sm2481@rit.edu

Ahmet Okutan
Rochester Institute of Technology
Rochester, NY, USA
axoeec@rit.edu

Mehdi Mirakhorli
Rochester Institute of Technology
Rochester, NY, USA
mxmvse@rit.edu

## ABSTRACT

The notion of *Attack Surface* refers to the critical points on the boundary of a software system which are accessible from outside or contain valuable content for attackers. The ability to identify attack surface components of software system has a significant role in effectiveness of vulnerability analysis approaches. Most prior works focus on vulnerability techniques that use an approximation of attack surfaces and there have not been many attempts to create a comprehensive list of attack surface components. Although limited number of studies have focused on attack surface analysis, they defined attack surface components based on project specific *hypotheses* to evaluate security risk of specific types of software applications. In this study, we leverage a qualitative analysis approach to empirically identify an extensive list of attack surface components. To this end, we conduct a Grounded Theory (GT) analysis on 1444 previously published vulnerability reports and weaknesses with a team of three software developers and security experts. We extract vulnerability information from two publicly available repositories: 1) *Common Vulnerabilities and Exposures (CVE)* and 2) *Common Weakness Enumeration (CWE)*. We ask three key questions: *where* the attacks come from, *what* they target, and *how* they emerge, and to help answer these questions we define three core categories for attack surface components: *Entry points*, *Targets*, and *Mechanisms*. We extract attack surface concepts related to each category from collected vulnerability information using the GT analysis and provide a comprehensive categorization that represents attack surface components of software systems from various perspectives. The paper introduces 254 new attack surface components that did not exist in the literature. The comparison of the proposed attack surface model with prior works indicates that only 6.7% of the identified *Code* level attack surface components are studied before.

## 1 INTRODUCTION

With thousands of new vulnerabilities being discovered every year, software security is more increasingly becoming a day-to-day concern for organizations across the world. Software security practitioners use a wide range of security analysis techniques to improve the confidentiality, integrity, and availability of software systems. These techniques often directly or indirectly rely on understanding application's *attack surfaces* —a set of points on the boundary of a software system, where an attacker can try to enter, cause an effect on, or extract data from [13, 18].

*Attack surface analysis*—the process of identifying applications' attack surface components (a.k.a points) plays a key role in numerous methods for security risk analysis [11, 13, 14, 18–20], vulnerability detection [4, 15–17, 28, 31, 35, 36, 38], and software testing [2]. Prior software vulnerability detection and testing approaches consider *Sink*, *Source*, *Entry point*, and *API/Function calls* as parts of the attack surface, however, these studies primarily focus on the analysis itself, rather than identifying attack surface components. There are a few studies that elaborate on the notion of the attack surface [11, 13, 14, 18–20, 28], consider entry points, exit points, channels, etc. as attack surface components, and test and validate them as existing theories [13, 14, 19, 20, 28]. They focus on limited-scope and example-based demonstration of attack surfaces of operating systems [13, 14, 18, 20] and web applications [11] using the attack surface metaphor. These studies show that applications with smaller attack surfaces are less vulnerable. While there has been a significant interest by practitioners and in the literature to study attack surface components of a given system, unfortunately, we lack a generic comprehensive guidance to support security research engineers in identifying attack surfaces of a given system. To the best of our knowledge, there is no prior research that takes a comprehensive approach to characterize and identify attack surface components in software systems.

In this paper we take a Grounded Theory (GT)-based approach [9, 10, 33] to characterize software attack surfaces and develop a comprehensive attack surface model to be reused by researchers and practitioners. Grounded theory is a qualitative approach that extracts theories from unstructured data and leads to discoveries directly supported by empirical evidence [7–10, 33]. We extract and analyze 810 vulnerability reports from Common Vulnerabilities and Exposures (CVE) data published by MITRE Corporation [26]. In addition, we analyze 634 entries in Common Weakness Enumeration (CWE) data, an extensive catalog of different types of software and hardware weaknesses describing root causes of vulnerabilities [27]. We leverage the Grounded Theory to identify high-level concepts which are related to software systems' attack surface from vulnerability reports and weaknesses, and use Straussian GT [5, 34] as a systematic inductive method for conducting qualitative research of identifying attack surface components. Our GT analysis starts by asking three key research questions:

- Research Question 1: Where are the critical *entry points* in a software system that are used by attackers to get in?
- Research Question 2: What assets or components in a software system are *targeted* by attackers?
- Research Question 3: How do attack surfaces emerge, and what types of *mechanisms* are utilized to reach the targets?

To answer these research questions we consider three core theories related to each research question, which are *Entry Points*, *Targets*, and *Mechanisms*, respectively. We extract the concepts related to each theory from gathered vulnerability data and define a generic attack surface model based on the emerged concepts. During the GT process we find that concepts related to each theory can be categorized in four major groups: software source code (*Code*), its executable (*Program*), the *System*, and the *Network* environment. Then, we identify attack surface components under each category and compare the proposed attack surface categorization model with the literature. The comparison results indicate that almost all attack surface components defined in the literature are covered by the proposed attack surface model, while prior works cover only a small portion of the concepts identified by our analysis. The result of quantitative comparison shows that in the best case only 50% and 20% of the *Network* and *Program* level mechanisms and 20% of the *Network* level entry points identified by this paper are covered in the literature. On average, only 6.7% of the studied *Code* level attack surface components are covered by previous works.

The remainder of this paper is organized as follows: Section 2 provides an overview of the methodology used in this empirical study, Section 3 presents the findings of our research, and Section 4 compares our results with those in the literature. Section 5 discusses the results, Section 6 describes verifiability and threats to validity, and Section 7 concludes with final remarks.

## 2 METHODOLOGY

Attack surface refers to the amount of code, functionality, and interfaces of a system exposed to attackers [14]. In this study, we rely on publicly available vulnerability repositories to identify common attack surface components in software systems. Vulnerability databases describe vulnerabilities using natural language and do not include technical data. In order to identify attack surface components, we use an approach based on the **Grounded Theory (GT)** [9, 10, 33].

The Straussian GT is preferred over the Classic Glaserian GT approach [7, 8], because the study is led by research questions and existing concepts in the literature are used during the analysis [5, 32, 34]. The Straussian grounded theory encompasses the following activities: (1) defining research questions, (2) theoretical sampling, (3) open coding, (4) constant comparisons, (5) memoing, (6) axial coding, and (7) selective coding. The GT process was performed by the authors who have had 5, 10, and 15 years of experience in software design, development, test, and maintenance, and 5, 5, and 10 years of experience in vulnerability analysis and secure by design, respectively. All three authors involved in all GT steps and the validation step that is discussed in Section 4. Figure 1 shows how the Straussian GT was applied. Over one year, the authors met weekly to discuss, merge and finalize the codes, concepts, and categories. All collected data, derived intermediate data that contains codes and concepts, and the final attack surface categories are shared with the research community through a public GitHub repository [1].

### 2.1 Research Questions

In Straussian grounded theory approach [5], researchers may define research questions upfront. While following the approach, we consider broad and open-ended research questions for detecting attack surface components. Getting inspired by the apparent attack surfaces of a house, research questions are defined based on the past experiences of the authors and concepts from the literature. For example, in a house, front and back doors, windows, garage door, climbable trees or tables can be entry points and the attacker would consider precious items in the house, such as safe box, as target. There might be some mechanisms in building a house such as emergency stairs that could make the house more vulnerable. Using the similarities of a software system and a house from the perspective of a cyberattacker, we identify the concepts for software applications to help define attack surface components. We focus on three research questions listed in Section 1 during the GT process and try to do coding in a way that can find theories from data to answer these questions.

### 2.2 Data Collection

Given the topic of interest of this study, we need access to software vulnerability reports, the description of these vulnerabilities, in-depth analysis of how they occurred, as well as vulnerable code snippets and their patches. Thus, we targeted open data resources that contain different types of vulnerabilities and vulnerable code snippets.

*2.2.1 Theoretical Sampling.* It is the data collection process that is based on the concepts derived from data [3, 5, 34]. In theoretical data sampling, unlike conventional approaches, all data are not collected at the beginning. Data collection and analysis is a circular process. Concepts that are identified in each cycle lead to more data collection until the saturation occurs [34].

*2.2.2 Data Sources.* We obtained vulnerability information from two publicly available vulnerability repositories:

- *Common Weakness Enumeration (CWE)*: CWE enumerates a list of common security weaknesses and categorizes them based on different views to help practitioners in securing their applications [27]. It provides a concise description of the weakness, common consequences, likelihood of exploitation, demonstrative examples, and reference to other resources. We reviewed all these information pieces to extract attack surface components.
- *Common Vulnerabilities and Exposures (CVE)*: The Mitre corporation CVE is an open platform to list publicly disclosed vulnerabilities [26]. We used the CVE list and additional information provided in the National Vulnerability Database (NVD) [29] to extract vulnerability meta-data.

We used issue tracking systems to obtain further discussions about CVEs, source code repositories to identify fixes for vulnerabilities, and further resources to extract other related information if existed. Figure 2 shows the information model of the vulnerability data collected. The summary of the data sources used:

①  **Retrieve vulnerabilities from MITRE and NVD**: We obtained vulnerability reports from MITRE CVE and NVD by consuming their public data feeds. Vulnerabilities disclosed in CVE are assigned a unique Identifier (CVE ID), a concise description, a list of affected software releases, and a list of references that can be used to obtain further details about the CVE, such as Issue Tracking Systems.

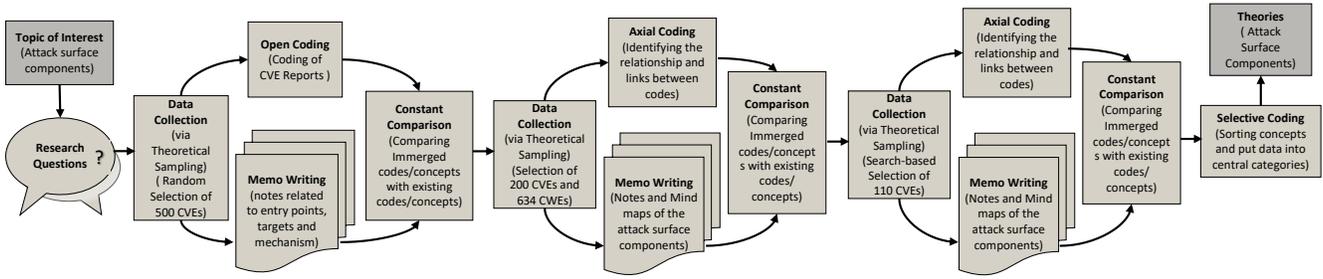

Figure 1: The Grounded Theory approach applied to our work

② **Obtain vulnerability details from issue tracking systems:** Although CVE reports provide information about different attributes of a vulnerability, they do not contain enough information to identify attack surface components at code level. Thus, we reviewed associated issue tracking systems for vulnerabilities that are related to open source projects. We leveraged the list of "references" to identify URLs to the corresponding bug entry of the issue tracking system and we read the developers' discussion about the problem, original code fragments, and their proposed solution(s).

③ **Gather patches from code repositories:** To retrieve patches that fixed vulnerabilities, we gathered the commits whose message explicitly mentioned the related bug id in the issue tracking systems or directly referred to the associated CVE. These patches often contained more information about the vulnerability, and the files that were affected, *i.e.*, modified, added or removed during the fix. Identifying patches helped us to identify entry points, targets and mechanisms at the code level.

④ **Collect vulnerability details from other references:** In addition to information that are provided in CVE website, we analyzed all the URLs that are provided as references for each vulnerability. These references include links to vulnerability reports, advisories or exploit information. These references provide more information for attack surface analysis of the vulnerability.

⑤ **Get vulnerability details from related CWEs** We identified related CWEs for each vulnerability. The CWEs helped us to understand the security issue of the CVE and extract attack surface components.

*2.2.3 Data Collection Process.* Number of vulnerabilities reported by NVD has increased from 6500 in 2016 to above 18000 in 2020 [30]. We focused on the vulnerabilities that have been reported during last five years (from 2016 to 2020 inclusive) to cover different types of weaknesses. For a more comprehensive attack surface analysis, we also collected data from *Introduced During Design/Implementation* views [21, 22] in *CWE* [27].

**First stage:** At the beginning, we selected 100 random vulnerabilities from each year (a total of 500) and collected their attributes.

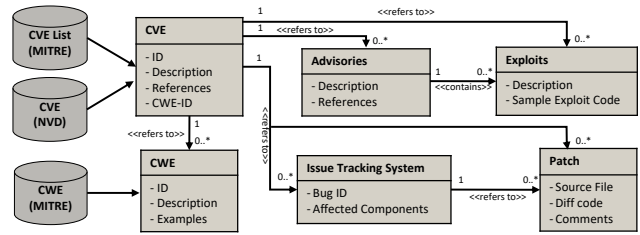

Figure 2: Information Model for the Collected Data

For some of these vulnerabilities only a CVE description was available (without any patch or advisory info), therefore, we were able to collect limited amount of information from the descriptions of such vulnerabilities. During the first stage of data collection, we performed coding process and extracted initial concepts.

**Second stage:** During this stage, we randomly selected 200 CVEs from 2016-2020 but omitted the areas that were theoretically saturated [3] during the first stage of the analysis. Therefore, we didn't do coding for the CVEs that were related to vulnerabilities such as *SQL injection*, *Buffer Overflow*, *Cross-site scripting*, and *Command injection* (36 CVEs), because reviewing more data related to these types of vulnerabilities no longer provided new theoretical insights about attack surface components at the end of the first stage. We noticed that CVEs collected during the first stage, covered limited number of CWE branches (70 CWE IDs), therefore we also collected data from *Introduced During Design and Introduced During Implementation* views [21, 22] in CWE (which totally contain 634 weaknesses after removing their common CWE IDs) to be more comprehensive. During our analysis in the second stage, we defined all components that can be part of an *Entry Point*, *Target* or *Mechanism*.

**Third stage:** In this stage, to identify new CVEs for emerged concepts that seemed incomplete and needed further analysis, we performed keyword-based selection of CVEs. We selected 110 CVEs related to these concepts. For example, for program architecture category we searched CVEs based on "Architecture", "Model", "Event Driven", "Master Slave", "Client Server", etc.

## 2.3 Open Coding

Open coding process analyzes collected data for each vulnerability and annotates them with codes (concepts) [5, 10]. We review the information gathered for each vulnerability (description, discussions, exploitation mechanism, and patches, *etc.*), annotate the information that is related to an *Entry Point*, *Target* or *Mechanism*, and assign code to the annotated key points. Code is a phrase that summarizes the key point in a descriptive way. Defined codes are assigned to the three general groups of attack surface components based on their relevance. Figure 3 shows how data was collected for CVE-2020-7248 as an example. In addition to the information available in MITRE website, we collected data from its security advisory (for exploit information), related GitHub repository (for source code), and CWE. The key points that are highlighted in red were extracted from both source code and descriptions and codes which are constructed based on key points assigned to the *Entry Point*, *Target*, and *Mechanism*:

- *Entry Point*: service request, ubus command, binary blob/json (input data);
- *Target*: stack memory, memory manipulation statement, memory copy statement, memcpy() function, buffer array;
- *Mechanism*: Serialization/Deserialization, JSON serialization;

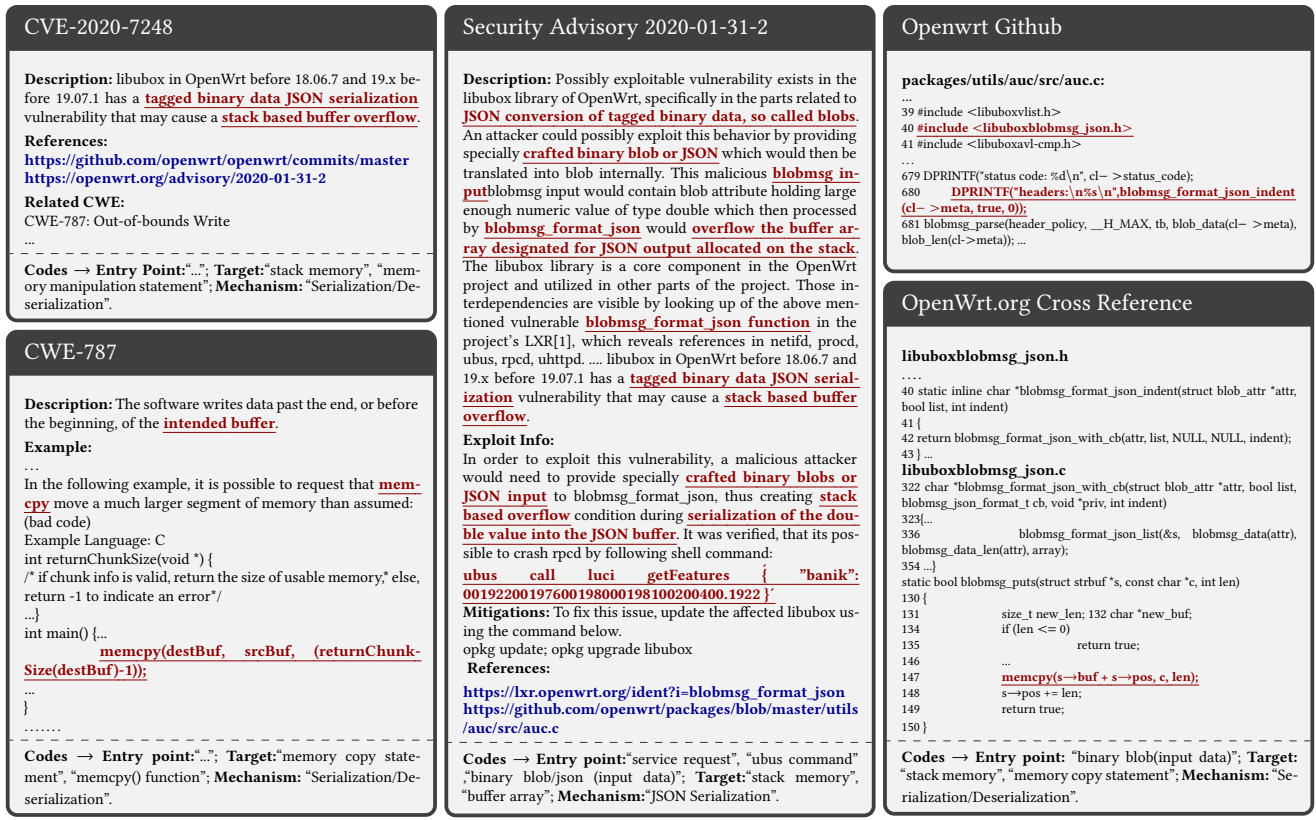

Figure 3: Open coding of data collected for CVE-2020-7248

Identified codes are constantly refined during the open coding process, leading to core categories and their associated concepts.

### 2.4 Constant Comparisons

We annotated vulnerability reports either by using the existing codes or creating new ones (if existing codes were not suitable for a newly analyzed vulnerability report). During the analysis of vulnerability reports, we also compare the existing concepts/categories against vulnerability reports to evolve categories and data interpretations. The overall goal of the open coding and constant comparison analysis is to identify the core concepts and categories related to the attack surface.

### 2.5 Memoing

Memos are notes, diagrams, or sketches that aid researchers to describe their preliminary ideas about properties and conceptual relationship between categories. After memoing, the researcher has stacks of memos in hand and puts them in an organized order by doing memo sorting [5]. Memoing is performed throughout the entire process of data coding and categorization. We used mind map diagrams to show the relationship between codes/concepts to identify core categories.

### 2.6 Axial Coding

During axial coding, we define new codes as a result of identifying new relationships [5, 34], and categorize the codes based on their relationship into higher level concepts. We perform axial coding for each research question separately. For instance, as shown in Figure 4, we categorize codes gathered from CVE-2020-7248 (Figure 3) and CVE-2016-9424 for the *Targets* into two higher level concepts: 1) *Application Code* (*Memory Manipulation*, *Array Access*) and 2) *Application Resource* (*Memory*). We define three higher level concepts for the *Entry Points*: 1) *Input data* (e.x. *BLOB/JSON*, *HTML Page*, and *HTML Tag*), 2) *Methods Receive Inputs*, and 3) *Service/Server Requests* (Figure 5). In this step, we also revisited the CVE instances in order to further refine their codes and attack surface components.

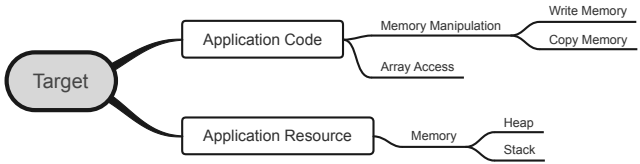

Figure 4: Mindmap example for *Targets*

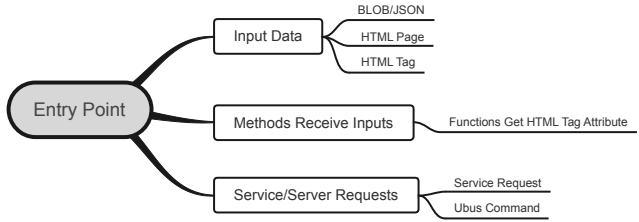

Figure 5: Mindmap example for *Entry Points*

*2.6.1 Data Analysis Instruments.* For the GT analysis, we developed a custom-built web-based tool to support our coding activities. This tool presents the information retrieved for each vulnerability report based on the research questions.

## 2.7 Selective Coding

In the last step of our analysis, we finalize the categorizations by sorting the defined concepts and associating them with the central branches, *i.e.*, the Program (P), Code (C), System (S) and Network (N) [34]. During this process, we integrate previously identified concepts and structure them into higher level of abstraction (theories) if needed. Selective coding re-organizes categories developed during axial coding.

## 3 RESULTS OF GROUNDED THEORY ANALYSIS

Based on the concepts emerged at the end of our analysis, we find that each core category *Entry Point*, *Target*, and *Mechanism* can be divided into four major groups, *i.e.*, Code (C), Program (P), System (S), and Network (N). From *Where* the attackers are entering into a system (entry points), *what* they are targeting for (targets) and *how* they are reaching the targets (mechanisms) are all related to the source *Code* of a software, its executable version named *Program*, the *System* that application is installed on, and the *Network* that the system is interacting with. The results of this qualitative study learned from reviewing CVEs and CWEs are presented based on

three research questions associated with *Entry Points*, *Targets* and *Mechanisms*:

## 3.1 Entry Points

Figure 6 shows the key categories we define for entry points based on the concepts identified at the end of our analysis to answer to Research Question #1. We define entry points based on four core categories:

*3.1.1 Code.* This category represents parts of source code that an attacker can leverage to enter a system. As shown in Figure 6 they are categorized into three sub categories:
①  **User Interface (UI)** defines components in the UI that can be used by attackers to enter a system. For example, an attacker can interact with an application through components in the graphical user interface (*Input Box*, *File Upload* [11], *RSS Feed* [11], *etc.*) or *Console*. ② **Methods/Directives** defines methods or directives that receive input. They can be parts of the code that directly receive input (*Direct Entry*) such as *Input Methods* [20, 28] that receive inputs directly from *User*, *Device*, or *File* or *Handlers* that handle different requests such as *OS Signal* (Interrupts) or web requests (REST API, Java Servlet). *Indirect Entry* covers parts of the source code that indirectly receive input by loading *Code*, reading *Indirect Inputs* (such as Environment Variable, System Attributes, *etc.*) or *User Created Resources*. ③ **Configuration File** category contains accessible configuration files of an application that can act as an indirect entry point for software application.

*3.1.2 Program.* This category considers an application as an executable and defines attack surface components related to that:
① **Components** refer to special software components that open the doors for attackers, such as application components which are designed during the design phase of software development. For example, *Plugin*, *Installer Components*, *Chatting Component*, and *Authentication/login* components are software components that can be considered as entry points at the design level. ② **Maintenance/Deployment** category covers any action that is performed during *Deployment* or *Maintenance* of programs that can open the doors for attackers. *Install*, *Configuration*, and *Update* operations are defined in this category. ③ **Direct Input** category covers data that is sent to the program. Application can receive *Direct Input* from *User*, *Device*, *Operating System (OS)*, or as *Messaging Object* from other components/applications (Intent in Android). It can also receive ④ **Indirect Input** by reading/loading *Environment Variables*, *DLL Files*, OMX buffer, *System Properties*, *Virtual Machine Properties*, and *Cookies* [11].

*3.1.3 System.* As a platform for running software applications, can provide entry points:
① **System input** contains both *Direct Input* and *Indirect Input*. *Direct Input* represents the requests that are sent to the system and is categorized into *Connection Requests* (SSH request), *OS Commands*, and *Service Requests*. *Indirect Inputs* represent the types of data imported by the system such as *Load Driver*. ② **Access Control** contains actions that may open the door for attackers. It contains *Local Access* to the system or *Improper Access Control*.

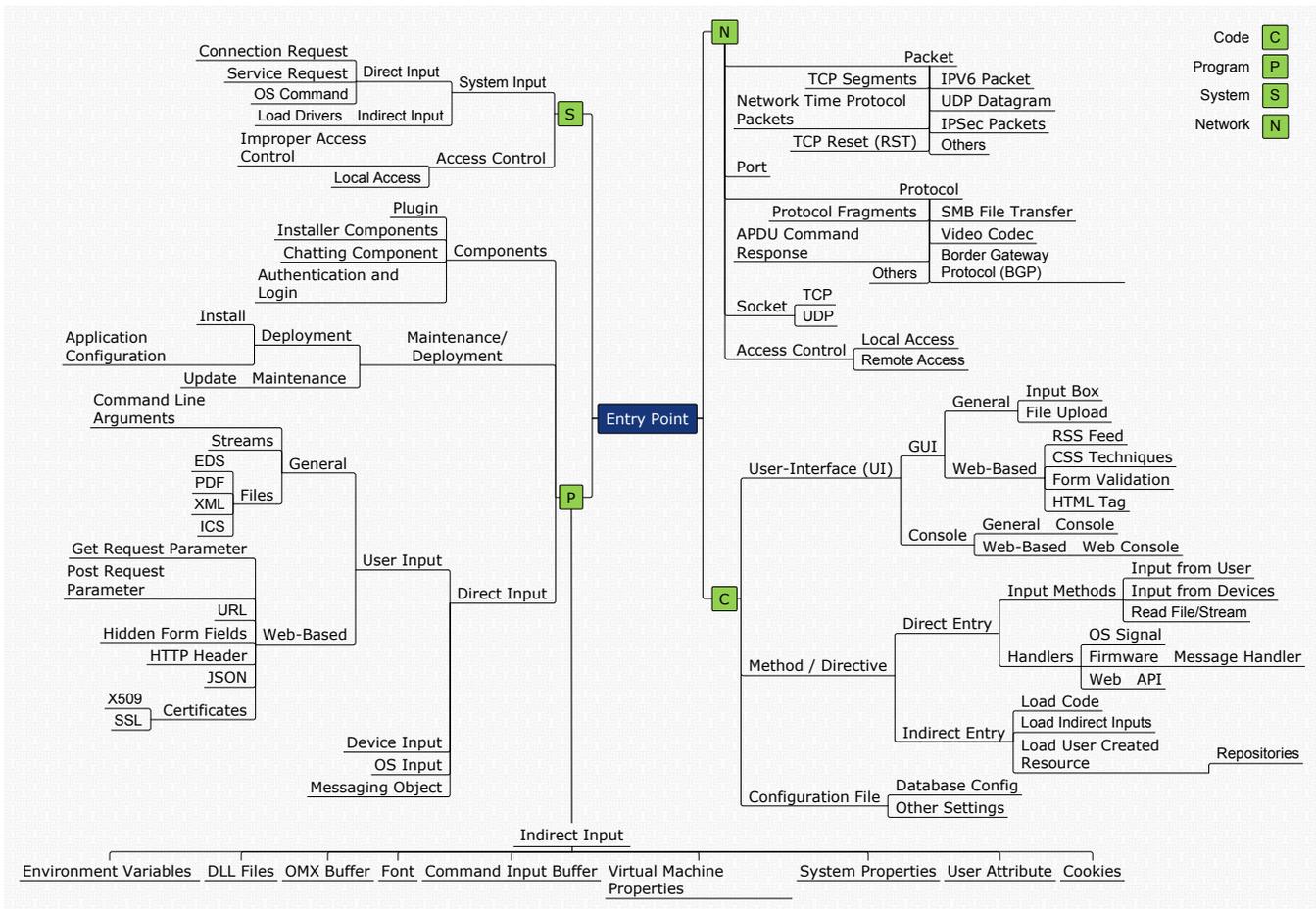

Figure 6: Identified *Entry Points* during attack surface analysis

*3.1.4 Network.* category contains the *Packet*, *Port*, *Protocol*, *Socket*, and *Access Control sub categories* : ① **Packet** represents the input data at the network level. ② **Socket** [14, 18, 20], ③ **Port**, and ④ **Protocol** [11, 14, 20] could provide entry points at the network level. ⑤ **Access Control** contains actions that open the doors for attackers such as *Local Access* to the network.

## 3.2 Targets

Figure 7 represents the categorization model created for *Targets* to answer to Research Question #2.

*3.2.1 Code.* This category defines source code related components that can be target of attacks. Attacker might try to access parts of source code to do malicious action. As shown in Figure 7 these components are categorized into two categories: ① **User Interface (UI)** category refers to components in the user interface that can be target of an attack. The analysis identified target components in this category such as *Validators* and *HTML/Webscript* that are related to *Web-Based* applications. ② **Method/Code Fragment** represents methods or other related parts of the code that can be target of attacks. As shown in Figure 7, parts of source code that handle requests (*Handlers*), execute *Commands* (*Database* or *OS* [28]), do *Memory Manipulation*, *Serialization/Deserialization*, *Reflection*, *Dangerous Operations* such as *Type Casting*, *Integer Operations*, *Encoding/Decoding*, *etc.*can be attractive targets for attackers. Besides that, code fragments such as *Exit points* [20, 28], *Critical Section*, some *Special Objects* such as *Gadget Classes*, *Cryptography Objects*, and *Path* are other targets at the code Level.

*3.2.2 Program.* The concepts under *Program* are categorized into two general categories: ① **Resource** contains resources allocated or used by the application such as *Memory. Stack*, *Heap*, *Cache*, *Shared Memory* [14] and other memory types that are allocated, used, or read by the application. ② **Data** covers important application data that are identified as target during GT analysis. This category considers application data from two perspectives: 1) *Data Resource* which represents the location where data is stored (*Database*, *File*, *etc.*) and 2) *Sensitive Information* that represents various kinds of important data that an attacker may look for. We found important files that can be target of attacks such as *Lock File*, *Log File*, *Certificate*, and *Keystore File. Credentials* such as *User*, *Used Service* (Notarization Service), and *Database* credentials, *Application*

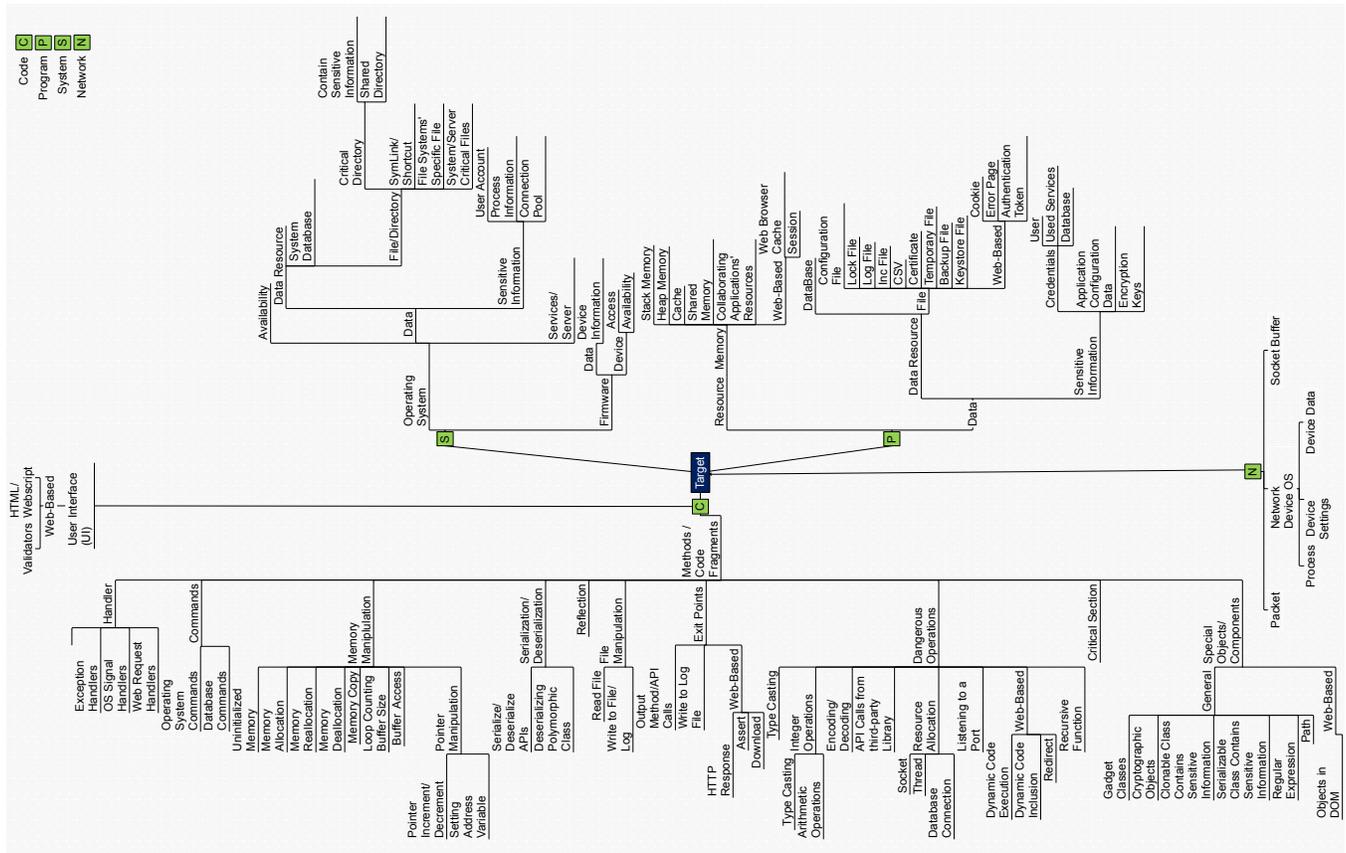

Figure 7: Identified *Targets* during attack surface analysis

*Configuration Data*, and *Encryption Keys* (ECDSA Secret, Master Key, *etc.*) are types of sensitive information identified during the coding process.

*3.2.3 System.* This category defines components in the OS or firmware that can be target of attacks. They can be directly accessed through attacks against the system or indirectly through attacks against software programs that use these components. These components are categorized into two major categories:

①**Operating System (OS)** contains OS and server related target components. They are categorized into different abstraction levels. *System Availability* covers actions that affect the availability of systems. For instance, a malicious System Reboot could interrupt a system and affect its availability. *System Data* categorizes different types of data in a system that can be target of attacks. It categorizes *System Data* based on the resource it is stored (*Data Resource*) and the type of the data (*Sensitive Information*). *Data Resources* can be a database on the OS (like Windows Registry [14, 20]) or an important *File/Directory* on the file system. *Critical Directory* may contain sensitive information (like etc/passwd in Linux), *Symlinks/Shortcuts* (like Unix Hard Link or Symbolic Link [18] and Windows Shortcut), *File System Specific Files* (like Data/Resource Fork of a File in HFS+ file system, Alternate Data Streams (ADS) in NTFS file system, *etc.*), and *System/Server Critical Files* (WSDL File in Web Server, Zone File in DNS Server, Node Catalogue in Distributed System, *etc.*). *User Account* information, *Process Information*, and *Connection Pool* are *Sensitive Information* in a system. *Services/Server* defines types of servers or services on a system that are usually target of attacks. For instance, SSL and NAS Servers are identified as targets in various vulnerability reports [23, 25].

②**Firmware.** category covers parts of firmware that contain *Device Information* or control *Device*.

*3.2.4 Network.* ①**Packets** and information on ②**Network Devices OS** such as *Process* (Routing Engine on Routers), *Device Setting*, and *Device Data*, and also ③**Socket Buffer** could be the target of attacks at the network level.

## 3.3 Mechanisms

This category answers the Research Question #3 by discussing mechanisms that are used at the source code, program, system, and network level that could lead to the emergence of vulnerable attack surfaces. Figure 8 represents the categorization model for *Mechanism*. We briefly summarize the mechanisms:

*3.3.1 Code.* Mechanisms used at the code level are categorized in three major categories:

①**User Interface (UI)** defines mechanisms used in the UI to open the doors for attackers. The concepts under this category are related to *Using Unsafe Techniques* in *Web-Based* applications such as *CSS Filters*, *RSS Feeds* [11], *Active Content* [11], and *Web Widget*.
②**Methods/Code Fragments** category discusses vulnerable mechanisms used during coding such as *Using Third-party Library*, *Serialization/Deserialization*, *Polymorphic Deserialization*, *Improper Security Check*, and *Authentication*, etc. *Improper Security Check* focuses on security mechanisms that are missed or implemented incorrectly such as *Improper Input Validation* [11], *Weak Encryption*, *Load Multiple Certificates* (like system and SSL certificate), and *Number of Requests*. *Authentication* category contains *Authentication Methods* (Password, Token-based, Certificate-based, *etc.*) and *Authentication Implementation* mechanisms such as *Location* (No Server Side Authentication) and *Insecure Implementation* (use == operator instead of === for Hash Comparison or unsafe info such as IP address for Authentication). ③ **Development Framework** category discusses vulnerable mechanisms which are related to the software development frameworks, like *Improper Configuration* (*Creating Debug Binary* or Improper *Encryption*) and *Installing Package Installers*.

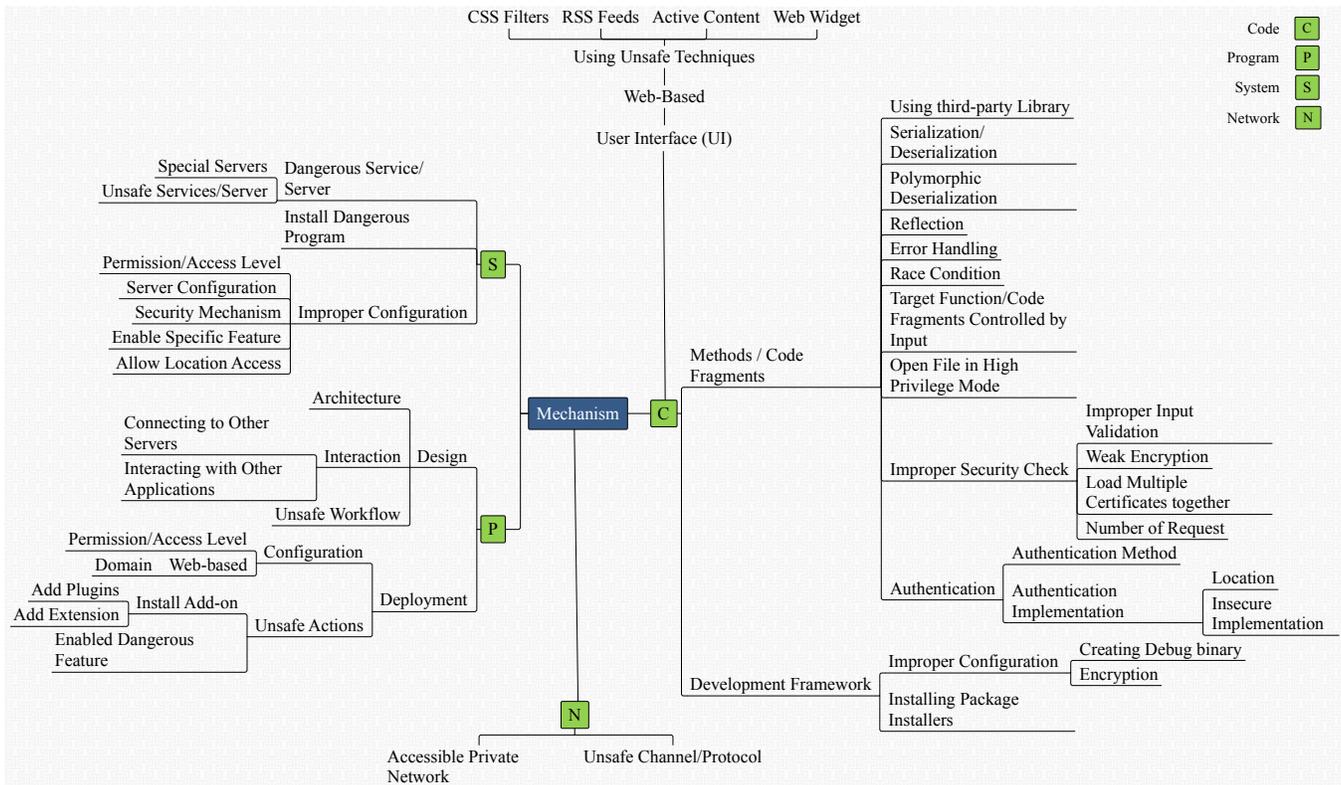

Figure 8: Identified *Mechanisms* during attack surface analysis

*3.3.2 Program.* This category categorizes the mechanism that are related to the design and deployment phase of a software system: ① **Design** category discusses mechanisms related to the design such as program *Architecture* (Client/Server, *etc.*), *Interaction* (connecting to *Other Servers* or *Other Applications*), and insecurely designed *Workflow*. ② **Deployment** category covers mechanisms during program deployment such as improper *Configuration* and *Unsafe Actions* that users can do on a system such as *Install Add-on* and *Enable Dangerous Features* [14].

*3.3.3 System Level.* category covers the following concepts:
① **Dangerous Services/Server** category refers to activating special *Servers* (e.x. SSL, NAS, and Mail) or installing/activating *Unsafe Services/Server* (like Unsafe FTP server) that can lead to emergence of attacks. ② **Dangerous Program** refers to installing special programs that can open the door for attackers (e.x. Android application that allows disabling/enabling WIFI to co-located apps [24]). ③ **Improper Configuration** refers to the configuration mechanisms that make the system vulnerable. These include *Permission/Access Level* (for File, Registry, *etc.*) [14, 18, 20], improper *Server Configuration*, *Security Mechanism* (Non-strict Security Mechanism in Firewall or Proxy), *Enable Specific Feature*, and *Allow Location Access*.

*3.3.4 Network Level.* mechanisms include ① **Accessible Private Network** and ② **Using Unsafe Channel/Protocol** [11, 14, 20] (using HTTP instead of HTTPS).

## 4 COMPARING TO RELATED WORK

The search strategy of our systematic literature review [39] consists of a manual search of five sources: the ACM Digital Library, IEEE Explore Library, ScienceDirect, Springer Link, and Google Scholar. Our inclusion criteria are as follows: the work is (i) a full paper; and (ii) focus on discussing software system attack surface.

Exclusion criteria are (i) position papers, short papers, keynotes, reviews, tutorial summaries, and panel discussions; (ii) not fully written in English; (iii) duplicated study; (iv) focused on attack surface outside the domain of software system; and (v) focused on attack surface of a specific type of system (e.x. IoT). We use the following search query: (Software OR Application) AND (Attack Surface OR Attack-Surface). From our manual search, we collected a total of 2,150 papers. We applied our inclusion and exclusion criteria through reading the paper's title, abstract, and keywords (if present), resulting in 30 papers. Then, in this round we applied the inclusion and exclusion criteria by reading the full papers, resulting in a remaining 8 papers. Some of the papers that were removed for further analysis, have misused the term "attack surface" (e.g. referring to software vulnerabilities). The remaining papers were carefully reviewed, to verify the extent to which the findings from our study were supported by the literature or were complementary. Limited studies have been proposed for identifying attack surface of a software system.

Howard et al [14] described an approach for measuring the relative attack surfaces of two systems with regard to certain dimensions. This study measures relative attack surface in three abstract and limited dimensions which are targets and enablers, channels and protocols, and access rights. Targets and enablers are resources that attackers can use such as process and data. For channels, they considered two types of channels: message passing and shared memory. They also considered account, privilege level and trust-relationship as access rights. They added three attack vectors to the 17 attack vectors that Howard [13] identified for windows system. Authors argue that their approach does not attempt to provide a comprehensive way of measuring the attack surfaces, but rather provides a relative way for comparing two versions of a system. This study is system level because they considered the services that run on a system.

Similarly, Manadhata and Wing in their work [18–20] proposed an attack surface metric to compare the security of two versions of a system to reason whether one is more secure than the other with respect to the attack surface metric. They proposed a model of a system and its environment using a state machine and considered any component that can be used to send/receive data to/from environment as an attack surface. They considered methods of the system, channels, and data items as resources. They used the model to identify the accessible subset of resources (in terms of access rights ) that contribute to the system's attack surface. They defined attackability of a resource based on its potential damage and the effort required to acquire access. However, similar to prior work they did not aim to define concrete and comprehensive attack surfaces. This study mainly focuses on metrics as approximations and shows the size of their metrics decreases when a vulnerability is patched.

Huemann et al. [11] defined the components of the attack surface for web applications. They proposed the attack surface of web application as a vector that has 22 dimensions that are categorized in 7 groups and considered weights for each components. They proposed Euclidean norm of the vector as an attack surface indicator.

Nuthan and Meneely [28] proposed function and file level attack surface metrics. They considered entry and exit points and also dangerous system calls as attack surface components. They provided static and static+dynamic call graphs and calculated the proximity and risky walk metrics based on the call graph. They proposed three proximity-based metrics which are proximity to entry points, exit points and dangerous points.

Theisen et al. [37] performed a systematic literature review on attack surface definitions. They categorized the attack surface into 6 categories which are methods, adversaries, flows, features, barriers, and reachable vulnerabilities. However, they mainly discussed granularity levels and concepts defined in previous works.

To evaluate the proposed attack surface categorization, we compare it with the attack surface components proposed in the literature. The comparison results (Table 1) show that the categorization provided by this paper covers all attack surface components introduced in the literature. The concepts which are missed in our categorization such as *Search* in *Program* and *RPC and Named Pipe* in *Network* level entry points are specific concepts that can be covered by the proposed core categories. The comparison results indicate that the proposed attack surface model differs from the previously introduced components in that it:

- Provides a comprehensive attack surface categorization that considers different aspects in *System*, *Network*, *Program*, and *Source* levels. Previous works mostly focused on defining attack surface components by low level concepts.
- Defines clear concepts that can be part of an attack surface. For instance, *Data Item* which is defined as an attack surface component in [20] is a vague concept. Based on the examples mentioned for it, our categorization clearly indicates that the data item could be program or system data.
- Provides comprehensive *Code* Level attack surface components. For example, previous studies considered I/O methods as *Entry Points* [20, 28] at the source level, however, we find that different *Handlers* and *Indirect Entry Points* can also be part of an attack surface. Nuthan and Meeneely [28] defined *System Calls* as *Dangerous Points*, however, we identify additional concepts as *Dangerous Points* such as *Type Casting*, *Integer Operations*, and *Encryption/Decryption*. We also define other code fragments such as *Serialization/Deserialization*, *Reflection*, *etc.* that can be target of attacks. Heumann et al. [11] define *URL Parameters* and Hidden Fields as input vectors, however, some other important input vectors such as *Post Request Parameter*, *HTTP Header*, and *Certificate* identified in our GT analysis are missed.

We compare the concepts defined by our GT analysis with the concepts defined in the literature. The comparison results are shown in Table 2. The results indicate that the literature covers a small percentage of *Code* level entry points, targets, and mechanisms, *i.e.*, 10%, 3.4%, and 10%, respectively. On average, at the *Code* level only 8 of the 119 concepts (6.7%) are covered by the literature. *Network* and *Program* level mechanisms, and *Network* level entry points are major categories that are covered in the literature with 50%, 20%, and 20%, respectively. In summary, the model proposed by this paper covers previously studied attack surface components and introduces 254 new concrete components that did not exist int the literature ($\sum_i (N_i - NL_i)$ in Table 2).

| | | Core Category | Low Level Concepts |
|---|---|---|---|
| Entry Points | C | UI | GUI: General (Input Box, File Upload [11]), Web-Based (RSS Feed [11], CSS Techniques, Form Validation, HTML Tag) Console: General (Console), Web-Based (Web Console) |
| | | Method [20]/ Directive | Direct Entry: Input Methods [11, 20, 28] (Input from User, Input from Devices, Read File/Stream), Handlers (OS Signal Handler, Firmware (Message Handler), Web (API)) Indirect Entry: (Load Code, Load Indirect Inputs, Load User Created Resources (Repositories)) |
| | | Config. File | Database Config, Other Settings |
| | | Components | Plugin, Installer Components, Chatting Component, Authentication and Login, Search [11] |
| | P | Maintenance/ Deployment | Deployment: Install, Application Configuration Maintenance: Update |
| | | Direct Input | User Input: General (Command Line Arguments, Streams, Files (EDS, PDF, XML, ICS )), Web-Based (Get Request Parameter [11], Post Request Parameter, URL, Hidden Form Fields [11], HTTP Header, JSON, Certificates (X509, SSL)) Device Input, OS Input, Messaging Object |
| | | Indirect Input | Environment Variables, DLL Files, OMX Buffer, Font, Command Input Buffer, Virtual Machine Properties, System Properties, User Attribute, Cookies [11, 20] |
| | S | System Input | Direct Input: (Connection Requests, Service Requests), Indirect Input: Load Drivers |
| | | Access Control [14, 20] | Improper Access Control, Local Access |
| | | Packet | IPV6 Packet, UDP Datagram, IPSec Packets, TCP Segments, TCP Reset (RST) Packet, Network Time Protocol Packets |
| | | Port | |
| | N | Protocol [11, 14, 20] | SMB File Transfer, Video Codec, Border Gateway Protocol (BGP), APDU Command Response, Protocol Fragments |
| | | Socket [14, 18, 19] | TCP [14, 18, 19], UDP [14, 18], RPC endpoint, named pipe [14, 19] |
| | | Access Control | Local Access, Remote Access |
| Mechanisms | C | UI | Web-Based: Using Unsafe Techniques (CSS Filters, RSS Feeds [11], Active Content [11], Web Widget) |
| | | Methods / Code Fragments | Using third-party Library, Serialization/Deserialization, Polymorphic deserialization, Reflection, Error Handling, Race Condition, Target Function/Code Fragments Controlled by Input, Open File in High Privilege Mode, Improper Security Check (Improper Input Validation [11], Weak Encryption, Load Multiple, Certificates together, Number of Request), Authentication ( Authentication Method, Authentication Implementation (Location, Insecure Implementation)) |
| | | Development Framework | Improper Configuration ( Creating Debug binary, Encryption), Installing Package Installers |
| | P | Design | Architecture, Interaction (Connecting to Other Servers, Interacting with Other Applications), Unsafe Workflow |
| | | Deployment | Configuration: (Permission/Access Level, Web-Based (Domain) [11]), Unsafe Actions: (Install Add-on (Add Plugins, Add Extension) , Enabled Dangerous Feature [14]) |
| | S | Dangerous Services/Server | Special Servers, Unsafe Services/Server |
| | | Install Dangerous Program | |
| | | Improper Configuration | Permission/Access Level [14, 18, 20], Server Configuration , Security Mechanism, Enable Specific Feature, Allow Location Access |
| | N | Accessible Private Network | |
| | | Unsafe Channel/ProtocoL [11] | |
| Targets | C | UI | Web-Based: Validators, HTML/Webscript |
| | | Methods / Code Fragments | Handler (Exception Handlers, OS Signal Handlers, Web Request Handlers), Commands (Operating System Commands [28] , Database Commands), Memory Manipulation (Uninitialized Memory, Memory Allocation, Memory Reallocation, Memory Deallocation, Memory Copy, Loop Counting Buffer Size, Buffer Access, Pointer Manipulation ( Pointer Increment/Decrement, Setting Address Variable)), Serialization/Deserialization (Serialize/Deserialize APIs, Deserializing Polymorphic Class), Reflection, File manipulation (Read File, Write to File/Log), Exit Points (Output Method/API Calls [20, 28], Write to Log File, Web-Based (HTTP Response, Assert, Download)), Dangerous operations (Type Casting, Integer Operations (Type Casting, Arithmetic Operations), Encoding/Decoding, API calls from third-party Library, Resource Allocation (Socket, Thread, Database Connection), Listening to a Port, Web-Based (Dynamic Code Execution, Dynamic Code Inclusion , Redirect), Recursive Function), Critical Section, Special Objects/Components (General (Gadget Classes, Cryptographic Objects, Clonable Class Contains Sensitive Information, Serializable Class Contains Sensitive Information, Regular Expression, Path), Web-Based (Objects in DOM)) |
| | P | Resource | Memory: Stack Memory, Heap Memory, Cache, Shared Memory [14], Collaborating Application Resources, Web-Based (Web Browser Cache, Session) |
| | | Data | Data Resource: Database , File (Configuration File, Lock File, Log File, Inc File, CSV, Certificate, Temporary File, Backup File, Keystore File, Web-Based (Cookie [11, 20], Error Page, Authentication Token)) Sensitive Information: Credentials (User , Used Service, Database), Application Configuration Data, Encryption Keys |
| | S | Operating System | Availability, Data (Data Resource (System Database [14, 20], File/Directory (Critical Directory (Contain Sensitive Information, Shared Directory), SymLink/Shortcut [18], File Systems' Specific File, System/Server Critical Files), Sensitive Information (User Account, Process Information, Connection Pool)), Services/Server |
| | | Firmware | Data: Device Information, Device: (Access, Availability) |
| | | Packet | |
| | N | Network Device OS | Process, Device Settings, Device Data |
| | | Socket Buffer | |

Table 1: Comparison of the concepts in the proposed attack surface model with the literature

## 5 DISCUSSION

Our analysis provides a very clear and comprehensive way to evaluate attack surface components at the *System*, *Network*, *Program*, and *Code* levels. While previous attack surface analysis works usually focus on specific types of systems like operating systems [13, 14, 18, 20] or web applications [11], this work defines a generic attack surface model that can be used to identify the attack surface components of different types of systems. Our work comprises attack surface components identified by earlier works [11, 14, 20, 28] and can be used to organize previously proposed concepts with more clear abstraction levels.

Proposed attack surface model represents critical parts of a software system that can be considered in various software analysis approaches. In *Requirement/Design* phase, the model can support threat modeling and identification of abuse cases initiated from entry points. In *Design*, software architects can use the catalog to measure and reduce attack surfaces, identify design decisions (e.x. connecting to other servers) or components (e.x plugins) that open new attack surfaces, increasing the risk of the system being attacked. Developers can use it during security code reviews to identify entry points and critical parts of code (target/mechanism) and verify the existence of security controls/patterns to secure those critical parts.

| Category | Level | N | NL | PL |
|---|---|---|---|---|
| Entry Points | C | 30 | 3 | 10 |
|  | P | 42 | 3 | 7.14 |
|  | S | 10 | 1 | 10 |
|  | N | 20 | 4 | 20 |
| Targets | C | 59 | 2 | 3.4 |
|  | P | 32 | 2 | 6.2 |
|  | S | 23 | 2 | 8.7 |
|  | N | 6 | 0 | 0 |
| Mechanisms | C | 30 | 3 | 10 |
|  | P | 15 | 3 | 20 |
|  | S | 10 | 1 | 10 |
|  | N | 2 | 1 | 50 |

Table 2: Quantitative comparison of the concepts in the proposed attack surface model with the literature. N shows the number of concepts identified in the model. NL and PL represent the number and percentage of concepts covered in literature, respectively

Testers can prioritize tests that examine the entry-points/targets. Pen-testers can use our catalogs as cheat-sheet to guide their testing activities.

## 6 VERIFIABILITY AND THREATS TO VALIDITY

While the aim of a GT study is to generate new theory, the verifiability of the theory can be inferred from the soundness of the research method. In this study, GT was strictly followed, each step was peer-reviewed and linked to the intermediary data to enable the reproducibility of findings. During the GT process, we implemented the triangulation concept to enhance the process validity [6]:

(1) *Data triangulation*: We collected data from a diverse set of CVEs which report real vulnerabilities from a variety of domains and also CWE that describes software security weaknesses. Additionally, as shown in Figure 3, for each CVE we looked at three interrelated data: vulnerability description from the product advisory, patches (source code), and exploit information.
(2) *Investigator triangulation*: Three authors worked together and performed the same GT steps. Over a period of one year, the authors met weekly, discussed, peer-reviewed, and finalized the code, memos, and emerged concepts.

The main limitation of our study is related to the GT method itself, because the validation phase of the GT process is challenging [12]. We mitigate this challenge partially by using literature as a source of validation. We evaluate the identified concepts by conducting a systematic literature review to explore how well these concepts fit to the previously studied software attack surfaces. The GT analysis includes an extensive manual analysis process and such manual analysis can be prone to biases. To help mitigate this threat, we followed the investigator triangulation method. Another limitation of this study is that the proposed model may reflect attack surfaces from recent vulnerability exposures, because the analysis covers CVEs between 2016 and 2020. To mitigate this threat partially, we included CWEs which are not time dependent as an additional data source.

## 7 CONCLUSIONS

Getting inspired from the similarity between the attack surfaces of a house/building and a software system, and asking three key questions (*Where* the attacks come from, *What* they target, and *How* they emerge), this paper develops a comprehensive attack surface model based on the *Entry Points* (Where), *Targets* (What), and *Mechanisms* (How) leveraged by cyberattackers. We follow a grounded theory-based approach to study attack surface components of software systems. Specifically, we focus on the software *Entry Points*, *Targets*, and *Mechanisms* to define the attack surface components in our model. We find that there are four major categories for each of these three branches, *i.e.*, *Code*, *Program*, *System*, and *Network*, and conduct a systematic literature review to verify to what extent previous studies corroborate with our findings. Preliminary results show that all attack surface components defined in the literature are covered by the proposed attack surface model, while prior works cover only a small portion of the concepts identified by our analysis. In the best case, the literature covers only 50% of *Network* level mechanisms, 20% of *Program* level mechanisms, and 20% of *Network* level entry points studied in this paper.